\documentclass[
aps
,floats,floatfix
]{revtex4}
\usepackage{graphicx}

\newcommand{\beq}{\begin{equation}}
\newcommand{\eeq}{\end{equation}}
\newcommand{\beqa}{\begin{eqnarray}}
\newcommand{\eeqa}{\end{eqnarray}}

\begin{document}

\title{Dispersion interactions in stratified anisotropic and optically active media\\ at all separations}
\author{Gregor Veble} 
\altaffiliation[Also at ]{Department of Physics, Faculty of Mathematics and Physics, Jadranska 19, SI-1000 Ljubljana, Slovenia, and Center for Applied Mathematics and Theoretical Physics,
University of Maribor, Maribor, Slovenia}
\affiliation{Faculty of Applied Sciences, University of Nova Gorica, Vipavska 13, P.O. Box 301, SI-5000 Nova Gorica, Slovenia}
 \author{Rudolf Podgornik}
 \altaffiliation[Also at ]{Institute of Biophysics, Medical Faculty, University of Ljubljana, and Department of Theoretical Physics, J. Stefan Institute, SI-1000 Ljubljana, Slovenia}
\affiliation{Department of Physics, Faculty of Mathematics and Physics, Jadranska 19, SI-1000 Ljubljana, Slovenia}

\date{\today}

\begin{abstract}
We propose a method to calculate dispersion interactions in a system composed of a one dimensional layering of finite thickness anisotropic and optically active slabs. The result is expressed within the algebra of $4 \times 4$ matrices and is demonstrated to be equivalent to the known limits of isotropic, nonretarded and uniaxial dispersion interactions. The method is also capable of handling dielectric media with smoothly varying anisotropy axes.
\end{abstract}

\maketitle

\section{Introduction}

Parsegian and Weiss \cite{Weiss} in their seminal paper of 1972 noted that dielectrically anisotropic materials generate van der Waals or dispersion torques that have the same origin as the more common and ubiquitous long range forces. Following this seminal contribution the effect of dielectric anisotropy in the dispersion interactions between two semiinfinite birefringent media has been studied theoretically in several contexts \cite{Enk,Shao} and has been given the most thorough analysis by Barash and coworkers \cite{Barash1, Barash2}. The outcome of these theoretical advances was that apart from the dispersion interactions that are present for any dielectric media, two birefringent slabs will also exhibit dispersion torques acting to align them. The magnitude of this torque is of course small but should nevertheless be detectable. Inspite of these theoretical advances there are as yet no experimental verifications of the theoretical predictions though there is plenty of activity in this research field \cite{Capasso}. It goes without saying that the dispersion torques should play a very important role in the context of micro and nano-mechanical actuators that could in principle transduce not only rectilinear motion but also rotation, or even couple and convert the two. 

In the present work we will formulate the theory of dispersion interactions in a system composed of a one dimensional layering of finite thickness  anisotropic dielectric slabs. In this way we generalize the already existing theory of dispersion interactions in a multilayer system of isotropic dielectric materials \cite{Podgornik2}. In this approach the secular determinant of the modes, which enters the free energy of the dispersion interactions, can be obtained as one of the elements of the transfer matrix of the EM field modes propagating through the system. The complete transfer matrix of the complete stratified system can be  decomposed into a product of two separate matrices, a diagonal propagator matrix and a symmetric discontinuity matrix. This method of matrix decomposition is extremely elegant and time saving for complicated one-dimensional stratifications and has been since used profitably in a variety of contexts \cite{profitably}.

The matrix approach introduced in the cited work has been formulated only for the case dispersion interactions between dielectrically isotropic media. For a complete solution of the anisotropic case one would need to extend it to solutions of a complete set of Maxwell equations with tensorial response functions. Indeed, we will show below that it can be extended and generalized to solve the dispersion interactions problem across stratified non-isotropic media. While the isotropic case can be formulated with the help of an algebra of 2 $\times$ 2 matrices, the general anisotropic case can only be implemented within an algebra of 4 $\times$ 4 matrices and is thus notoriously more difficult to handle, but still simpler than would follow from alternative approaches \cite{Barash1}. Nevertheless we are able to extract some informative limiting cases that show the power and book-keeping elegance of this approach.

\section{Formalism}

The propagation of electromagnetic waves in layered anisotropic media was described already by Berreman \cite{Berreman} who introduced a convenient and efficient formalism that allows for a transparent formulation of a notoriously extremely complicated problem \cite{Barash1}. In what follows we basically follow this approach. If $z$ is the direction perpendicular to the translational plane of symmetry of the problem, so that the dielectric slabs of anisotropic material are layered in this direction, the propagation of the single frequency electromagnetic field in such a system can be Fourier-decomposed in the $x-y$ plane perpendicular to the $z$ direction with wavenumber vector $\vec Q$ and described using the four dimensional vector of the field components
\beq
\psi(z; \vec Q)=\left[
\begin{array}{c}
E_x(z; \vec Q)\\
H_y(z; \vec Q)\\
E_y(z; \vec Q)\\
-H_x(z; \vec Q)
\end{array}
\right].
\eeq
The Fourier field components evolve as a function of the coordinate $z$ via a homogeneous system of linear differential equations that stem directly from the properly formulated Maxwell equations
\beq
\frac{\partial}{\partial z} \psi=\frac{i \omega}{c} {\bf D} \psi,
\eeq
where ${\bf D}={\bf D}(\underline\epsilon(\omega,z),\underline \mu(\omega,z), \underline \rho(\omega,z),\underline \rho^\prime (\omega,z),\vec Q,\omega)$ is a $4\times4$ complex matrix that depends on the frequency and $z$ coordinate dependent $3\times3$ tensors of the electric and magnetic permeabilities $\underline\epsilon$ and $\underline \mu$, the $3\times3$ tensors of optical activity $\underline \rho$, $\underline \rho^\prime$, the twodimensional vector $\vec Q$ in the $x-y$ plane that gives the wavenumber in the plane of symmetry and the chosen angular frequency $\omega$. 

The details on how to construct ${\bf D}$ are given in \cite{Berreman}, with the main results repeated for consistency here.
A $6\times6$ matrix ${\bf M}$ is constructed from the material response tensors as
\beq
{\bf M}=\left[\begin{array}{cc}
\underline \epsilon & \underline \rho \\
\underline \rho^\prime & \underline \mu \\
\end{array}\right].
\eeq

An auxiliary matrix ${\bf a}$ is then calculated as
\beqa
a_{3, 1} & = & [M_{6, 1} M_{3, 6}  - M_{3, 1} M_{6, 6}]/d, \\
a_{3, 2} & = & [(M_{6, 2}  - c Q/\omega)M_{3, 6}  - M_{3, 2} M_{6, 6}]/d, \\
a_{3, 4} & = & [M_{6, 4} M_{3, 6}  - M_{3, 4} M_{6, 6}]/d, \\
a_{3, 5} & = & [M_{6, 5} M_{3, 6}  - (M_{3, 5}  + c Q/\omega)M_{6, 6}]/d,\\
a_{6, 1} & = & [M_{6, 3} M_{3, 1}  - M_{3, 3} M_{6, 1}]/d,\\
a_{6, 2} & = & [M_{6, 3} M_{3, 2}  - M_{3, 3} (M_{6, 2}  - c Q/\omega)]/d,\\
a_{6, 4} & = & [M_{6, 3} M_{3, 4}  - M_{3, 3} M_{6, 4}]/d,\\
a_{6, 5} & = & [M_{6, 3} (M_{3, 5}  + c Q/\omega) - M_{3, 3} M_{6, 5}]/d,
\eeqa
with
\beq
d = M_{3, 3} M_{6, 6}  - M_{3, 6} M_{6, 3}.
\eeq
Without any loss of generality the wavenumber $\vec Q$ is assumed to lie in the $x$ direction and we have to rotate the coordinate frame accordingly. The elements of a $4\times4$ matrix ${\bf S}$ are then defined as
\beqa
S_{1, 1}  & = & M_{1, 1}  + M_{1, 3} a_{3, 1}  + M_{1, 6} a_{6, 1}, \\
S_{1, 2}  & = & M_{1, 2}  + M_{1, 3} a_{3, 2}  + M_{1, 6} a_{6, 2}, \\
S_{1, 3}  & = & M_{1, 4}  + M_{1, 3} a_{3, 4}  + M_{1, 6} a_{6, 4}, \\
S_{1, 4}  & = & M_{1, 5}  + M_{1, 3} a_{3, 5}  + M_{1, 6} a_{6, 5}, \\
S_{2, 1}  & = & M_{2, 1}  + M_{2, 3} a_{3, 1}  + (M_{2, 6}  - c Q/\omega)a_{6, 1}, \\
S_{2, 2}  & = & M_{2, 2}  + M_{2, 3} a_{3, 2}  + (M_{2, 6}  - c Q/\omega)a_{6, 2}, \\
S_{2, 3}  & = & M_{2, 4}  + M_{2, 3} a_{3, 4}  + (M_{2, 6}  - c Q/\omega)a_{6, 4}, \\
S_{2, 4}  & = & M_{2, 5}  + M_{2, 3} a_{3, 5}  + (M_{2, 6}  - c Q/\omega)a_{6, 5}, \\
S_{3, 1}  & = & M_{4, 1}  + M_{4, 3} a_{3, 1}  + M_{4, 6} a_{6, 1}, \\
S_{3, 2}  & = & M_{4, 2}  + M_{4, 3} a_{3, 2}  + M_{4, 6} a_{6, 2}, \\
S_{3, 3}  & = & M_{4, 4}  + M_{4, 3} a_{3, 4}  + M_{4, 6} a_{6, 4}, \\
S_{3, 4}  & = & M_{4, 5}  + M_{4, 3} a_{3, 5}  + M_{4, 6} a_{6, 5}, \\
S_{4, 1}  & = & M_{5, 1}  + (M_{5, 3}  + c Q/\omega)a_{3, 1}  + M_{5, 6} a_{6, 1}, \\
S_{4, 2}  & = & M_{5, 2}  + (M_{5, 3}  + c Q/\omega)a_{3, 2}  + M_{5, 6} a_{6, 2}, \\
S_{4, 3}  & = & M_{5, 4}  + (M_{5, 3}  + c Q/\omega)a_{3, 4}  + M_{5, 6} a_{6, 4}, \\
S_{4, 4}  & = & M_{5, 5}  + (M_{5, 3}  + c Q/\omega)a_{3, 5}  + M_{5, 6} a_{6, 5},
\eeqa
so that the sought for matrix {\bf D} is then given as
\beq
{\bf D}={\bf H}\cdot {\bf J} \cdot {\bf S} \cdot {\bf H}^{-1}
\eeq
with a matrix corresponding to a permutation of components
\beq
{\bf H}=
\left[\begin{array}{cccc}
1 & 0 & 0 & 0\\
0 & 0 & 0 & 1\\
0 & 1 &  0 & 0\\
0 & 0 & 1 & 0\\
\end{array}\right]
\eeq
and a transformation of the form
\beq
{\bf J}=
\left[\begin{array}{cccc}
0 & 0 & 0 & 1\\
0 & 0 & -1 & 0\\
0 & -1 &  0 & 0\\
1 & 0 & 0 & 0\\
\end{array}\right].
\eeq

If the matrix ${\bf D}$ is constant within an interval $[z,z+h]$, then the propagation of EM field modes within that interval can be written in terms of a matrix exponent
\beq
\psi(z+h)={\bf P}(h)\psi(z),\ \ {\bf P}(h)=\exp\left[\frac{i \omega}{c} {\bf D} ~h \right].
\eeq
Here the matrix ${\bf P}$ is a constant within that interval but does depend on the length of the interval, $h$.
A matrix exponent of the form above can be calculated by performing a diagonalisation of ${\bf D}$
\beq
{\bf P}(h)={\bf R} ~{\bf K}(h)~ {\bf R}^{-1},\ \ K_{jj}=\exp\left(\frac{i \omega}{c} d_j ~h \right),
\label{eq:exponential}
\eeq
where ${\bf K}$ is a diagonal matrix, $d_j$ are the eigenvalues of the matrix ${\bf D}$ and the rotation matrices ${\bf R}$ and ${\bf R}^{-1}$ correspond to the left and right eigenvectors of ${\bf D}$. The matrix ${\bf R}^{-1}$ performs a mapping from the vector $\psi$ of the electromagnetic field components into the vector of two left and two right propagating plane wave components that we denote by $\eta$.

Propagation of EM waves across a stratified set of $N-1$ layers of finite width, that are bounded by semi-infinite half-spaces from both directions, can be described by taking a product
\beq
\tilde {\bf P}={\bf P}_N(h_N)~{\bf P}_{N-1} (h_{N-1})\ldots {\bf P}_1(h_1)~{\bf P}_0(h_0) 
\eeq
or
\beq
\tilde {\bf P}={{\bf R}_N {\bf K}_N(h_n)} {\bf R}_N^{-1}~{\bf R}_{N-1} {\bf K}_{N-1}(h_{n-1}) {\bf R}_{N-1}^{-1}
\ldots {\bf R}_0 {{\bf K}_0(h_0) {\bf R}_0^{-1}}.
\label{tyuer}
\eeq
where the indices $0$ and $N$ correspond to the left and right half-spaces. This is indeed very similar to the case treated in \cite{Podgornik} but in that case the dielectric media are assumed to be isotropic.

A propagator of the form Eq. \ref{tyuer} maps the fields from a point that is at some arbitrary depth $h_0$ in the left half-space to a point that is at another arbitrary depth $h_N$ in the right half-space. The propagators ${\bf P}_0$ and ${\bf P}_N$, however, do nothing more than rotate the phases of the incoming and outgoing plane waves. 

Furthermore, the propagator $\tilde {\bf P}$ maps the components of the electric and magnetic fields, whereas it is the mapping of plane wave amplitudes that is of interest for dispersion interactions. A mapping of plane wave amplitudes from one side of the stratified system of slabs to the other is obtained by the truncation of $\tilde {\bf P}$ on both the left and the right sides of the expression to obtain
\beq
{\bf P}= {\bf R}_N^{-1}~{\bf R}_{N-1} {\bf K}_{N-1}(h_{N-1}) {\bf R}_{N-1}^{-1}
\ldots {\bf R}_{1} {\bf K}_{1}(h_1) {\bf R}_{1}^{-1}~{\bf R}_0.
\eeq
In order to propagate the vector of plane wave amplitudes from the external boundary of the leftmost layer to the boundary of the rightmost one than has
\beq
\eta^{(N)}={\bf P} \eta^{(0)}. \label{eq:waveprop}
\eeq
This propagator does not depend on the arbitrary thicknesses $h_0$ and $h_N$ anymore. The only information that the bounding half-spaces contribute to the propagator is their respective transformations from the electromagnetic fields into plane waves as given by the matrices ${\bf R}_0$, ${\bf R}_N^{-1}$.

With these expressions it is now possible to tackle the problem of dispersion interactions in such media in analogy with \cite{Podgornik}. The dispersion interaction free energy $F$ per area $A$ is given as
\beq
F/A=kT {\sum_{n=0}^{\infty}}^{\prime} \int \frac{d^2 \vec Q}{(2 \pi)^2} \ln {C(i\xi_n)}.
\label{eq:freeen}
\eeq 
The prime in the summation over the Matsubara thermal frequencies denotes that the $n=0$ term is to be taken with the weight of $1/2$. Here $C(\omega)$ is an expression whose zeros give the eigenfrequencies of bound states.  It is evaluated as a function of the imaginary Matsubara frequencies $\xi_n=2\pi n k T/\hbar$. In order to evaluate the above interaction free energy we first need to calculate  $C(\omega)$ in terms of the propagator equation Eq. \ref{eq:waveprop}.

We define a state as bound if it is exponentially decreasing to both $-\infty$ and to $+\infty$. Let the four components of the plane wave amplitude vector $\eta$ for each layer be sorted with respect to their real parts such that the {exponentially increasing solutions} have {indices $1,2$} and the {decreasing ones} have {indices $3,4$}. In order for the state to be bound, the exponentially decreasing components of $\eta_0$ and the exponentially increasing components of $\eta_N$ in the corresponding half-space must be equal to 0. This gives for the wave propagation equation \ref{eq:waveprop} corresponding to bound states the form
\beq
\left[
\begin{array}{c}
{\eta_1^{(N)}=0}\\
{\eta_2^{(N)}=0}\\
\eta_3^{(N)}\\
\eta_4^{(N)}
\end{array}
\right]=\left[\begin{array}{cc}
\\
\ {{\bf P}_I} &\  {\bf P}_{II}\\
\\
\ {\bf P}_{III} &\  {\bf P}_{IV}\\
{\ }
\end{array}\right]
\left[
\begin{array}{c}
\eta_1^{(0)}\\
\eta_2^{(0)}\\
{\eta_3^{(0)}=0}\\
{\eta_4^{(0)}=0}
\end{array}
\right]
\eeq
that can be clearly reduced to a $2\times2$ homogeneous system of equations
\beq
{{\bf P}_I}\left[
\begin{array}{c}
\eta_1^{(0)}\\
\eta_2^{(0)}\\
\end{array}
\right]=0.
\eeq
This system of equations allows us to identify the bound state condition that gives their eigenfrequencies as
\beq
C={\det {\bf P}_I=0}. \label{eq:det}
\eeq
When evaluating $C$ at imaginary frequencies in equation (\ref{eq:freeen}), the determinant $\det {\bf P}_I$ will in general have a complex value. We may take its absolute value as the appropriate contribution to the free energy, since the values of $C$ for the opposite wavenumbers $\vec Q$ and $-\vec Q$ are complex conjugate and therefore the imaginary values of their logarithms cancel.

As the free energy expression is indeterminate up to an additive constant, this means that the bound state condition is undetermined up to a constant scaling factor. It is natural to choose such a scaling that further splitting of the half-spaces into sublayers with equal properties does not alter the results. This will also ensure that the dispersion interactions only depend on the spatial variations of the electromagnetic response tensors and not on their homogeneous values. This is acheived by shifting the eigenvalues of the ${\bf D}$ matrix by 
\beq
{\tilde d_i=d_i -\bar d},\ \ {\bar d=(d_1+d_2)/2} \label{eq:eigenshift}
\eeq
and use these when evaluating the propagators ${\bf K}_i$.

There is still an additional freedom of scaling factors due to the eigenvector normalisations. Since the matrix ${\bf D}$ is in general not hermitian, its left and right eigenvectors are not complex conjugate and there is no unique way to normalise them individually. Care must therefore be taken when comparing systems whose terminating half spaces are being rotated with respect to each other. A common reference point (such as the case of large separation) must be calculated in order to be able to determine the free energy shift due to eigenvector rotations.

One particular reference point for any system can be a corresponding system in which all the layers between the two limiting half-spaces are substituted by vacuum, and the limit of the distance between the two half-spaces (the vacuum distance) is taken  to infinity. This configuration presents no interaction between the two half-spaces, and therefore the calculated free energy contribution of this configuration can be considered as the sought shift. For this corresponding system, the matrix ${\bf P}_c$ is given by
\beq
{\bf P}_c={\bf R}_N^{-1}~{\bf P}_\infty~{\bf R}_0, \label{eq:reference}
\eeq
where ${\bf P}_\infty$ corresponds to the vacuum propagator for the electromagnetic waves in the limit of long distances. Calculating it using the expression (\ref{eq:exponential}) and the shift (\ref{eq:eigenshift}) gives
\beq
{\bf P}_\infty=
\left[\begin{array}{cccc}
\frac{1}{2} & -\frac{\tilde \rho}{2}& 0 & 0\\
-\frac{1}{2\tilde \rho} & \frac{1}{2} & 0 & 0\\
0 & 0 &  \frac{1}{2} &-\frac{1}{2\tilde \rho} \\
0 & 0 &-\frac{\tilde \rho}{2}& \frac{1}{2}\\
\end{array}\right],\ \tilde \rho=\sqrt{1+\frac{Q^2 c^2}{\xi ^2}}.
\eeq
The free energy contribution for this corresponding system is then calculated via (\ref{eq:det}) and and (\ref{eq:freeen}). In this way one can determine the free energy shift due to eigenvector rotations and properly normalize the dispersion interaction free energy. This concludes our derivation of the general formalism. In what follows we will consider a few interesting limiting cases.

\section{Isotropic case}

When the response tensors are isotropic, the general approach introduced above should reduce to the formalism introduced in \cite{Podgornik}. Let us assume $\underline \epsilon=\epsilon \underline I$,  $\underline \mu=\mu \underline I$ and  $\underline \rho=\underline \rho^\prime=0$. The matrix ${\bf D}$ for this case reduces to the simple form
\beq
 {\bf D}=
\left[\begin{array}{cccc}
0 & \mu+ \frac{Q^2 c^2}{\xi ^2\epsilon}& 0 & 0\\
\epsilon & 0 & 0 & 0\\
0 & 0 &  0 & \mu\\
0 & 0 & \epsilon+  \frac{Q^2 c^2}{\xi ^2\mu}& 0\\
\end{array}\right]
\eeq
where $\xi=-i\omega$ is the imaginary frequency. The eigenvalues of such a matrix are doubly degenerate and come in pairs of $d_i=\pm \nu$, where 
\beq
\nu=\sqrt{\epsilon \mu +\frac{c^2 Q^2}{\xi^2}},
\eeq
such that ${\bf D}={\bf R}\ {\rm diag} \left\{d_i\right\} {\bf R}^{-1}$ can be written in the form
\beq
{\bf D}=\left[\begin{array}{cccc}
-\nu & \nu & 0 & 0\\
\epsilon & \epsilon & 0 & 0\\
0 & 0 &  -\mu & \mu\\
0 & 0 & \nu& \nu\\
\end{array}\right]
\left[\begin{array}{cccc}
-\nu & 0 & 0 & 0\\
0 & \nu & 0 & 0\\
0 & 0 &  -\nu & 0\\
0 & 0 & 0 & \nu\\
\end{array}\right]
\left[\begin{array}{cccc}
-\frac{1}{2\nu} & \frac{1}{2\epsilon} & 0 & 0\\
\frac{1}{2 \nu} & \frac{1}{2\epsilon} & 0 & 0\\
0 & 0 & -\frac{1}{2\mu} & \frac{1}{2\nu}\\
0 & 0 & \frac{1}{2\mu} & \frac{1}{2\nu}\\
\end{array}\right].
\eeq
The matrices in the above expression are block diagonal, so the four dimensional problem decomposes into a pair of two dimensional ones that correspond to the TE and TM modes of the electromagnetic propagation. Let us elaborate this further.

After shifting the eigenvalues in order to remove the exponential divergence, the propagator ${\bf P}$ is then given as
\beq
{\bf P}={\bf R}
\left[\begin{array}{cccc}
1 & 0 & 0 & 0\\
0 & \exp(-2 \rho) & 0 & 0\\
0 & 0 &  1 & 0\\
0 & 0 & 0 & \exp(-2 \rho)\\
\end{array}\right]
{\bf R}^{-1},
\eeq
where $\rho=\frac{\xi}{c} \nu$ in order to conform with the notation in \cite{Podgornik}. The transition of eigenmodes between two subsequent layers is given by the matrix
\beq
{\bf R}_{i+1}^{-1} {\bf R}_i=\frac{1}{2}
\left[\begin{array}{cccc}
\frac{\epsilon_i}{\epsilon_{i+1}}+\frac{\rho_i}{\rho_{i+1}} & \frac{\epsilon_i}{\epsilon_{i+1}}-\frac{\rho_i}{\rho_{i+1}} & 0 & 0\\
\frac{\epsilon_i}{\epsilon_{i+1}}-\frac{\rho_i}{\rho_{i+1}} &\frac{\epsilon_i}{\epsilon_{i+1}}+\frac{\rho_i}{\rho_{i+1}} & 0 & 0\\
0 & 0 &  \frac{\mu_i}{\mu_{i+1}}+\frac{\rho_i}{\rho_{i+1}}& -\frac{\mu_i}{\mu_{i+1}}+\frac{\rho_i}{\rho_{i+1}}\\
0 & 0 &   -\frac{\mu_i}{\mu_{i+1}}+\frac{\rho_i}{\rho_{i+1}} &  \frac{\mu_i}{\mu_{i+1}}+\frac{\rho_i}{\rho_{i+1}}\\
\end{array}\right].
\eeq
If we further split the above matrix into separate TE and TM diagonal blocks, and apply two arbitrary scaling factors due to the eigenvector normalisations by dividing the diagonal matrices with their diagonal terms, we obtain
\beq
G_i=\left[\begin{array}{cc}
1 & -\Delta_i \\
-\Delta_i & 1
\end{array}
\right]
\eeq
with
\beq
\Delta_i=\frac{\epsilon_{i+1} \rho_i-\epsilon_i \rho_{i+1}}{\epsilon_{i+1} \rho_i+\epsilon_i \rho_{i+1}}
\eeq
for the TM and
\beq
\Delta_i=\frac{\rho_{i+1} \mu_i-\rho_i \mu_{i+1}}{\rho_{i+1} \mu_i+\rho_i \mu_{i+1}}
\eeq
for the TE block. By also denoting 
\beq
T_i=\left[\begin{array}{cc}
1 & 0\\
0 & \exp(-2 \rho_i)
\end{array}
\right]
\eeq
we can obtain the free energy contributions as the $\{1,1\}$ matrix element of a matrix product,
\beq
C_{\rm TE,\ TM}=\left[G_N T_N \ldots G_1 T_1 G_0\right]_{\{11\}},
\eeq
which agrees exactly with the result derived in \cite{Podgornik}.

In the full $4\times4$ formulation there does not exist a single common scaling factor for the matrix ${\bf R}$ which would exactly map it to the $2\times2$ formalism of \cite{Podgornik}. Scaling factors are, however, only related to the shift in the free energy, which is itself undetermined up to a constant.

\section{The single slab anisotropic case}

The nonretared case of two semi-infinite anisotropic dielectric media separated by a slab of a third anisotropic medium, where for all the media one of  the anisotropy axes coincides with the $z$ axis, was derived in \cite{Weiss}. This result can be summarized as follows.

Let the dielectric tensors for the first half space, the slab and the second half space be given by $\underline \epsilon_1$, $\underline \epsilon_2$ and $\underline \epsilon_3$, respectively. Each tensor can be written in the form
\beq
\underline \epsilon_i=
\left[\begin{array}{ccc}
\epsilon_{i,1} & 0 & 0\\
0 & \epsilon_{i,2} & 0 \\
0 & 0 & \epsilon_{i,3}
\end{array}\right].
\eeq
Each of the materials is rotated around the $z$ axis such that
\beq
\underline {\tilde \epsilon}_i=O_i \underline \epsilon_i O_i^{-1},
\eeq
where 
\beq
O_i=\left[\begin{array}{ccc}
\cos (\theta_i)& -\sin(\theta_i) & 0\\
\sin(\theta_i) & \cos (\theta_i)& 0 \\
0 & 0 & 1
\end{array}\right].
\eeq
From a given deilectric tensor the quantity
\beq
G(\underline \epsilon_i, \alpha)^2=\frac{\epsilon_{i,1}}{\epsilon_{i,3}}
 + \frac{\epsilon_{i,2}- \epsilon_{i,1}}{\epsilon_{i,3}} \sin^2(\alpha)
\eeq
may be calculated. From this
\beq
\beta(\underline { \epsilon}_i, \theta_i, \varphi) = Q~G(\underline \epsilon_i, \theta_i - \varphi)
\eeq
is defined, where $\varphi$ is the direction of the plane parallel wavevector for which the free energy contribution is being evaluated. Further
\beq
a=\frac{\epsilon_{1,3} \beta(\underline {\epsilon}_1, \theta_1, \varphi)}{
\epsilon_{2,3} \beta(\underline {\epsilon}_2, \theta_2, \varphi)}
\eeq
and 
\beq
b=\frac{\epsilon_{3,3} \beta(\underline {\epsilon}_3, \theta_3, \varphi)}{
\epsilon_{2,3} \beta(\underline {\epsilon}_2, \theta_2, \varphi)}
\eeq
are introduced, from which 
\beq
\Delta^2=\frac{(a-1)(b-1)}{(a+1)(b+1)}
\eeq
is calculated. The free energy contribution for a fixed size of the plane parallel wave vector $Q$ and its angle $\varphi$ is given by
\beq
C_P(\underline {\epsilon}_1,\underline {\epsilon}_2,\underline {\epsilon}_3, \theta_1, \theta_2, \theta_3, Q,\varphi,L)=1-\Delta^2 \exp\left(-2 QG(\underline {\epsilon}_2,\theta_2-\varphi) L \right),
\eeq
and the total free energy then follows as
\beq
F/A=\frac{kT}{4 \pi^2} {\sum_{n=0}^{\infty}}^\prime  \int_0^{2 \pi} d\varphi \int_0^{\infty}  \ln C_P~Q ~dQ. 
\eeq

While we were not able to demonstrate analytical equivalence of our approach with the above result, numerical computations nevertheless agree {\sl perfectly}, up to a constant shift. As already explained, the constant free energy shift in our formulation is a result of arbitrary eigenvector normalization, and the results must be compared to some chosen point, which in this case is the infinite separation of the two materials. Note that this constant energy shift makes no contribution to the interaction forces, given by the derivative of the free energy.

\begin{figure}
\centerline{\includegraphics[width=8cm]{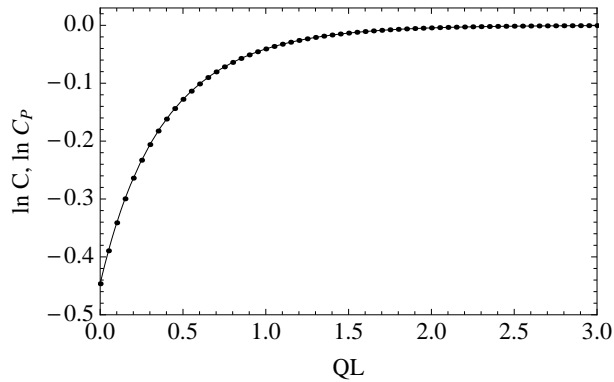}}
\caption{Anisotropic free energy contributions as a function of separation $L$ for the nonretarded case (line) compared to our approach (dots). Both free energy expressions coincide numerically to any desired accuracy. See text for details. \label{fig:parsegian}}
\end{figure}

For comparison purposes, we choose the free energy contributions from our method, $C$, in the limit of inifinite $c$, and the corresponding nonretarded expression, $C_P$, calculated for the dielectric response tensors at some given imaginary frequency. These are arbitrarily chosen to be 
$\epsilon_{1,1}=5$, $\epsilon_{1,2}=10$,  $\epsilon_{1,3}=15$, 
$\epsilon_{2,1}=1$, $\epsilon_{2,2}=3$,  $\epsilon_{2,3}=2$, 
$\epsilon_{3,1}=18$, $\epsilon_{3,2}=12$,  $\epsilon_{3,3}=6$. The second medium is rotated by the angle $\theta_2=2$ with respect to the first one and the third one by $\theta_3=1$ with respect to the first one. The angle for the wavevector was chosen as $\varphi=3$. The free energy contributions for both calculations as a function of distance are given in Fig. \ref{fig:parsegian}. For our method, the large distance free energy result was subtracted from all the results in order to eliminate the shift in the free energy contribution. The match is exact for this set of parameters as well as for any other one might choose. This strongly supports the general conclusion that the non-retarded form of our calculation coincides exactly with the calculation derived in \cite{Weiss}. 

A calculation for the fully retarded dispersion interaction of the same configuration is given in \cite{Barash1}, with the limitation that the materials are uniaxial and that the axis of anisotropy lies in the symmetry plane of the problem, with the medium between the two slabs being isotropic. The result is convoluted and lengthy, and will thus not be repeated here (for details see \cite{Barash1,Barash2,erratum}). Using the same notation as before, the arbitrarily chosen values are
$\epsilon_{1,1}=5$, $\epsilon_{1,2}=10$,  $\epsilon_{1,3}=10$, 
$\epsilon_{2,1}=2$, $\epsilon_{2,2}=2$,  $\epsilon_{2,3}=2$, 
$\epsilon_{3,1}=6$, $\epsilon_{3,2}=3$,  $\epsilon_{3,3}=3$. The third medium is rotated by the angle 
$\theta_3=1$ with respect to the first one. The angle for the wavevector is chosen as $\varphi=2$. The value of the imaginary frequency is $\xi=1 Qc$. The comparison of the free energy contribution of the retarded calculation (denoted as $\ln C_B$) and our results are given in figure \ref{fig:parsegian}. The results again match exactly for this set of parameters as well as for any other one might choose.
\begin{figure}
\centerline{\includegraphics[width=8cm]{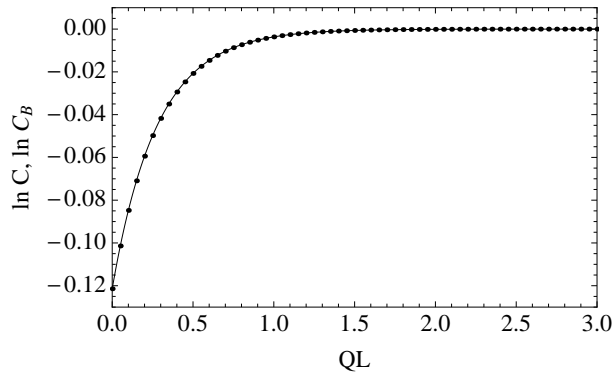}}
\caption{Anisotropic free energy contributions as a function of separation $L$ for the retarded case (line) compared to our approach (dots). See text for details. \label{fig:barash}}
\end{figure}

This concludes our numerical proof that all the limits with which one can meaningfully compare our result are reached exactly within the numerical computation. Again, we have not been able to establish these correspondences analytically. The exact expressions are unfortunately extremely convoluted and complicated, thus precluding any meaningful and straightforward simplification.

\section{Smooth Anisotropic Profiles}

There is one obviously restrictive condition that is built into the Lifshitz theory of dispersion interactions, which underlyes also the approach presented in this work: this is the assumed steplike change in the dielectric permeability at the interfaces of bodies that 
interact across spatially homogeneous media. Several extensions of the theory have relaxed the condition of steplike 
interfaces as well as formulated the dispersion interactions in a non-uniform dielectric media \cite{nonuniform} where the response tensors are assumed to vary smoothly and continuously in the $z$ direction. Since this extension of the dispersion interactions theory was built on the reformulation of the Lifshitz theory based on the algebra of 2$\times$2 matrices, we assume that the case of continuously varying anisotropic response tensors could be obtained by a proper generalization of the method derived in this work.

Indeed it turns out that our method can be applied to the case of smoothly varying response tensors as functions of the $z$ variable. As a model, we take an anisotropic dielectric material with the zero frequency tensor components 
$\epsilon_{1,1}=5$, $\epsilon_{2,2}=10$,  $\epsilon_{3,3}=20$ in its diagonal reference frame. In the coordinate frame where the $z$ direction is perpendicular to the translational plane of symmetry, the material is taken to be rotated first by an angle of $1$ around the $x$ axis, and then by an angle of $2$ around the $y$ axis. The corresponding rotated dielectric response tensor is denoted by $\underline \epsilon_l$. Then, the tensor $\underline \epsilon_r(\delta)$ is obtained by further rotating the material by an additional angle $\delta$ around the $z$ axis. The model for the dielectric response is then 
\beq
\underline \epsilon(z)=\left\{
\begin{array}{ccc}
z<0 & ; & \underline \epsilon_l\\
0\leq z \leq z_0 & ;  & \cos^2 \left(\frac{\pi z}{2 z_0}\right)\underline \epsilon_l +\sin^2 \left(\frac{\pi z}{2 z_0}\right)\underline \epsilon_r(\delta) \\
z_0<z & ; & \underline \epsilon_r(\delta) \\
\end{array} 
\right. \label{eq:smoothprof}
\eeq
This profile can, for example, be considered as a model for a grain boundary between two grains of an anisotropic material \cite{adrian}.

We model the continuous dielectric material as a composition of $N_l=100$ thinly stratified discrete layers of thickness $z_0/N_l$, where each layer is assigned a constant value of the dielectric response given by equation (\ref{eq:smoothprof}) with $z$ taken at the middle of the layer. We then apply our general theory to write down the dispersion interaction free energy between the two outermost semi-infinite dielectric regions with thinly stratified region in between. The discretization can be done with any value of $N_l$ and in the limit approaches the continuum result as was proven exactly for the non-retarded isotropic case \cite{equivalence}. 

In order to compute the free energy (\ref{eq:freeen}) an assumption needs to be made on both the frequency $\xi$ as well as wavevector $Q$ dependence of the dielectric response. For clarity of the model, we assume that the response is constant in both $\xi$ and $Q$ until their limiting values of $\xi_0$ and $Q_0$, at which point the dielectric response becomes $1$ throughout the material. In the calculations, we choose $Q_0=1/z_0$ and $\xi_0=c/z_0$.

\begin{figure}
\centerline{\includegraphics[width=8cm]{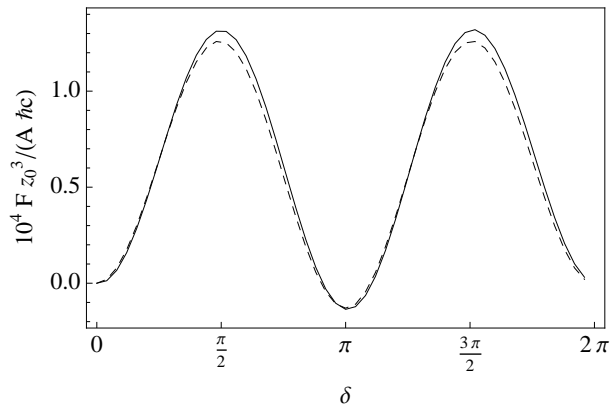}}
\caption{Anisotropic free energy density as a function of the rotation angle $\delta$. Solid line denotes the infinite separation free energy normalization, the dashed line represents the slowly varying dielectric response normalization. See text for details. \label{fig:smooth}}
\end{figure}

As has already been mentioned, the free energy calculations require a proper reference point due to the arbitrariness of the eigenvector normalizations. Two reference points were taken for calculations at each rotation angle $\delta$. The first one was the large empty space separation of the outermost layers as summed up by equation (\ref{eq:reference}), and the other was by taking a very slowly varying profile of the same shape as given in (\ref{eq:smoothprof}) but with $z_0=30$ instead of $1$, with the relatively low number chosen for numerical stability reasons. The free energy density as a function of the rotation angle $\delta$ is shown for both normalization approaches, Fig. \ref{fig:smooth}. The small residual difference between the two methods of eigenvector normalization  can be attributed to the non-infinite separations in the slowly varying case. 

Our general method for evaluation of dispersion interactions in complicated one-dimensional geometries involving non-isotropic dielectric materials obviously predicts the existence of torques, {\sl i.e.} derivatives of the free energy with respect to $\delta$, between the two outermost homogeneous semi-infinite materials. It would thus certainly add to the overall energy balance equation of the formation of a grain boundary between two grains of anisotropic crystals, a research direction we will not pursue in this paper.

\section{Conclusions}

We devised a method that allows us to numerically calculate dispersion interactions in a system composed of an arbitrary number of anisotropic and optically active finite dielectric slabs. The result is expressed in the formalism of $4 \times 4$ matrices which, for the most general case, requires a numerical approach. We showed that the results analytically reduce to known formulations in the general isotropic case, and that, for a single slab, it gives the same numerical result in the nonretarded and uniaxial limits. One of the main strengths of the method is that it is also capable of handling smoothly varying media, where the medium is represented as a discretized set of small thickness slabs.

\section{Acknowledgements}

This work has been supported by the European Commission under contract No. NMP3-CT-2005-013862 (INCEMS) and by the Slovenian Research Agency under contract  No.  J1-0908 (Active media nanoactuators with dispersion forces).

\end{document}